\documentclass[12pt]{article}
\usepackage[numbers]{natbib}
\usepackage{amsmath, amssymb, amsthm}
\usepackage{hyperref}
\usepackage{graphicx}
\usepackage{setspace}
\usepackage{geometry}
\usepackage{xcolor}
\usepackage{outlines}
\usepackage{caption}
\usepackage{tikz}
\usetikzlibrary{arrows, positioning}
\usepackage{pgfplots}
\usetikzlibrary{patterns, fillbetween}
\pgfplotsset{compat=1.18}
\usepackage{mathtools}
\usepackage[normalem]{ulem}
\usepackage{tfrupee}

\newcommand{\p}{\mathcal{P}}

\title{Preconditions necessary for trade}
\title{Preconditions for the existence of economic processes}
\title{Economy (or lack thereof) in certain alternative universes}
\title{Axiomatic aspects of Economy}
\title{Physical Aspects of the Classical Economic Theory}
\title{How the economy depends on a law of Physics}
\title{Economy as an Outcome of Conserved Quantities}
\title{Axiomatic Incompleteness of the Classical Economic Theories}
\title{Necessity of Conserved Quantities for Axiomatic Completeness of Classical Economic Theories}

\author{Sidharth Gat \\
sgat@ncsu.edu $\vert$ sidharth.gat2000@gmail.com \\
ORCID: 0000-0002-1244-5643}
\date{September 2025}

\begin{document}

\maketitle

\begin{abstract}

{
Many economic theories have been introduced over the course of history to articulate our understanding of the economy. Classical theories by Adam Smith 
and David Ricardo's Comparative Advantage 
have been foundational for the last century's work. Improvements have been achieved over time, incorporating insights from many disparate fields of study: contemporary frameworks like Behavioural Economics
, and Information Economics
, incorporate psychological insights and deviation from rational decision-making, and insights from network theory and how the information flow affects the market behaviour, respectively. In this paper, I motivate the necessity of incorporating insights from Physics, and also show that trade as a phenomenon described by the comparative advantage theory cannot exist without the law of conservation of Energy, and incorporating this law leads to axiomatic completeness of the theory. Further, I also argue that while the economy is not a zero-sum game in terms of wealth, it does require at least one associated zero-sum parameter for trade and economy as phenomena to exist.}

\end{abstract}

{
\section{Introduction}

Many economic theories have been introduced over the course of history to articulate our understanding of the economy. Classical theories by Adam Smith \cite{smith1776} and David Ricardo's Comparative Advantage \cite{comparative-advantage} have been foundational for the last century's work. Many improvements have been achieved over time, incorporating insights about the world from many disparate fields of study. Contemporary frameworks like Behavioural economics \cite{Behavioral_economics_pioneered_by_Daniel_Kahneman_and_Amos_Tversky}, and Information Economics \cite{Information_economics_pioneered_b_George_Akerlof_Michael_Spence_and_Joseph_Stiglitz_explores_the_role_of_information_asymmetry_in_economic_transactions}, incorporate psychological insights and deviation from rational decision-making, and insights from network theory and how the information flow affects the market behaviour. In this paper, I motivate the necessity of incorporating insights from Physics. The Laws of Physics govern everything that happens around us. They control the unfolding of our world. They determine how exactly a particle would interact with its surroundings, and the surroundings with the particle, at a microscopic level. An immeasurable number of these interactions is what gives rise to everything we can see around us in the world and everything that happens within us. The flickering of stars, geological activities, climatic systems, the evolution of life from one form to another, the behaviour of these species, the differential behaviour of individuals within a species, and even something as personal as my choice of food and work. Due to the vast number of microscopic interactions, computing the future (though fundamentally probabilistic) by computing all the laws of physics for each of these interactions is impossible to keep track of. We might not necessarily understand why a particular phenomenon is the way it is. It, however, remains true that they obey and are even given rise to by the most fundamental laws of our nature. This paper explains how a law of physics governs the economy, and more specifically, forms the basis for trade decisions. 

In Sec. \ref{Phenomenology} I start with a theoretic description of the economic process derived from the classical economic theories (\cite{smith1776} and \cite{ricardo1817}). In Sec. \ref{Formalism}, I develop the necessary mathematical formalism and prove the necessity of conserved quantities by stating the aforementioned classical economic theory into a set of axioms, and showing the mathematical contradictions that arise as a consequence of excluding the law of conservation of energy as one of the axioms. In sec. \ref{sec:corollaries:Energy is not wealth though.} and \ref{sec:corollaries:Energy is the zero-sum parameter, not wealth.}, I elaborate connection between trade to physical quantities like \textit{Energy} expenditure, introducing the concept of the \textit{would-have-been consumed Energy}. In Sec. \ref{sec:corollaries:Energy is the zero-sum parameter, not wealth.}, I also argue that while the economy is not a zero-sum game in terms of wealth, it does require at least one zero-sum parameter (to be used interchangeably with conserved quantity\footnote{\textit{Zero-sum} is a common terminology in Economics. In this paper, I append it with the word \textit{parameter} to distinguish it from the reference of wealth being zero-sum, and explicitly refer to a variable that is zero-sum: as in {\textit{zero-sum parameter}---something whose net quantity stays the in the system, or, equivalently, one individual's gain [of this quantity] is another's loss [of this quantity]. A \textit{conserved quantity} is a common phrasing in Physics, which means the same thing. I will use either interchangeably in this paper.}) associated with the process of labour and trade. 
}like \textit{Energy}. Lastly, in sec. \ref{sec:Errenous estimates}, I also discuss how errors in the estimation of individuals lead to imperfect decisions and deviations from the ideal trade and jobs execution choices.
}

{\color{gray}

\color{gray} 
}

{
\section{Essence of Comparative Advantage Theory}
\label{Phenomenology}
}

Note: Related sentences are superscripted with same number at the end, throught.

{




Humans used to do a lot more of the work needed for their sustenance by themselves \cite{soo2011gains}\footnote{While this section provides theoretic description of the economic dynamics as proposed by \cite{ricardo1817}, many validating pieces of evidence are presented in the original work by \citet{smith1776} and \citet{ricardo1817}, as well as \citet{ridley2010rational}}. The food the people ate, they cooked themselves; for the raw material needed, they farmed it themselves; even for the fuel required to cook food, families scavenged it themselves; they built their own houses and collected the material needed to build them. The diversity of work an individual engaged in was immense: people didn't specialise. None dedicated their days to a singular kind of activity. There wasn't anyone dedicating their day to providing the masses (including themselves) with drinking water, hoping that other work necessary for their sustenance could be outsourced. Why did specialisation arise, then? Was this mode of being disadvantageous? The answer is Yes \cite{comparative-advantage}. 

This mode of life was (and is\textsuperscript{4}) suboptimal, due to the differences in individuals' abilities \cite{ridley2010rational}. An individual good at working on a farm, perhaps due to physical vigour (or even due to farming acuity†\textsuperscript{6}), also had to spend part of his day on other things. Things he isn’t, and someone in his neighbourhood is, good at (relatively). The gain, let’s say, of time or energy savings (because he is an adept farmer, for whatever reasons, and variation in time or energy spent to do the same job could be taken as a measure of variation in adeptness), while working on a farm would be nullified by the loss of the same by engaging in other chores over a day—or, more accurately, balanced by similar gain by those in his neighbourhood.

As time progressed, individuals learned about the strengths of their peers and those of their own. Specialisation took root (a bit more than earlier). If it costs me more Energy than the guy next door to cook the same amount of food, I would rather have him cook for both of us. If he finds I do a better job at making shelters than him, in similar terms, he would rather have me build houses for both of us. For we find it makes both of us richer (even though it can do so to different degrees\textsuperscript{5}). We specialise, I, a constructor, and he, a cook. We exchange labour (implicitly if not explicitly). The trade sets in.

Moreover, specialisation, rather than being a once-occurred-in-history phenomenon, is a perpetual, quasi-static process\textsuperscript{4}. It grows deeper and deeper as time progresses. My trade mate might further segregate his business; he might choose to only cook rather than farm, collect wood for fuel, make utensils (in whatever form they existed then), harvest water, and, only then, cook food to its end form. He would do so if he found that he could increase the net worth of his net produce (easy to measure and thus understand, with the currency system rather than the barter system). 

Now, if two players on the market happen to be good at the same thing, the more efficient one would earn the larger margin just by virtue of being efficient (assuming he can accommodate the demand and his efficiency isn’t a negative function of the amount he produces). Even if their products/services are priced equally, the one who spent less money in the process of bringing them into existence earns a greater margin. Or, equivalently, the one who has expended less energy.

The competition between two players is bound to take the prices down, but the efficient one remains profitable when the other can barely survive. The lowest price you could get it at is the price most efficient producers can afford to sell it at. So, the lower bound of prices is dictated by how much it costs the producer [of goods/services] to produce it. Second, the prices can’t be taken too high, even in the absence of competitors, for the consumers might find they are wealthier producing it themselves (or going without it). In other words, as long as there are customers, there are competitors. If one charges \rupee80 or lower per meal, I’m better off buying it. It saves me time and energy, and I can be more productive—and thus, profitable (Of course, this is a difficult problem, and the most optimum point along the price axis that maximises my return might not be at \rupee80). If instead, I’m being charged \rupee500 per meal, then, to the best of my estimation, I would be wealthier cooking food myself. So, the lower bound of the prices is dictated by whether the buyer finds it profitable to outsource (as opposed to producing by himself).

This is precisely how one decides when to loosen up his grip on a paper note and trade it for something else\textsuperscript{2}. The goal is to maximise the overall margin. It is to arrive at the best solution to the optimisation problem designed to maximise the margin. The Input to this function is—everything—you can think of, literally. 'Everything' because individuals practice their tasks and improve their efficiency, attaining local maxima on the efficiency manifold for the given task. Individuals also learn, which helps them spot maxima beyond the local maxima, not only allowing them to do the tasks radically differently, but also invent new tasks altogether, which might be a non-trivial union or segregation of previously existing tasks.}

{
\citet{ridley2010rational} puts together a vast number of studies showing evidence for progressive division of labour throughout history. The mechanistic description I gave here is derived from one of David Ricardo's (\citet{ricardo1817}). I will use it to put the dynamics on an axiomatic footing in \ref{Formalism}, where I present a concrete demonstration of the need for a conserved quantity, provided in Sec \ref{Formalism}, and the stage for an extended discussion of the above Ridardian description, introducing the concepts of would-have-been-consumed energy,

}

{
\section{Formalism}
\label{Formalism}
}

\begin{figure}[h]
\centering
\begin{tikzpicture}[
SIR/.style={rectangle, draw=red!60, fill=red!5, very thick, minimum size=5mm},]

\node[SIR]    (j)                              {$j_{\vec{\epsilon}}$};
\node[SIR]    (xyz)       [below=of j] {$x$, $y$, $z$ ...};
\node[SIR]    (price)       [below=of xyz] {$N(P_j - P_0)$};
\node[SIR]    (max)       [below=of price] {wants to maximise};

\draw[->, very thick] (j.south)  to node[right] {needs outcomes of} (xyz.north);
\draw[->, very thick] (xyz.south)  to node[right] {can be obtained by expending a limited amount of energy/money available} (price.north);
\draw[->, very thick] (price.south) to node[right] {maximising this allows maximising of the energy/money available} (max.north);
\end{tikzpicture}
\caption{A player's decision tree showing player $j$ (characterised by $\vec{\epsilon}$), with ultimate goal of maximising his profits or minimising systems \textit{would-have-been-consumed energy}, and using decision of trading of executing $x$, $y$, $z ...$, as intermediary means.}
\label{individual}
\end{figure}
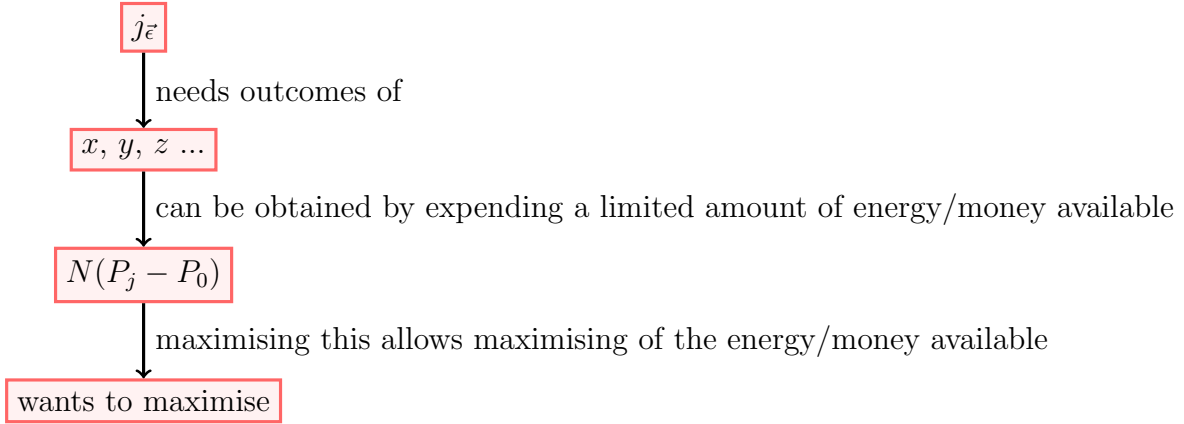

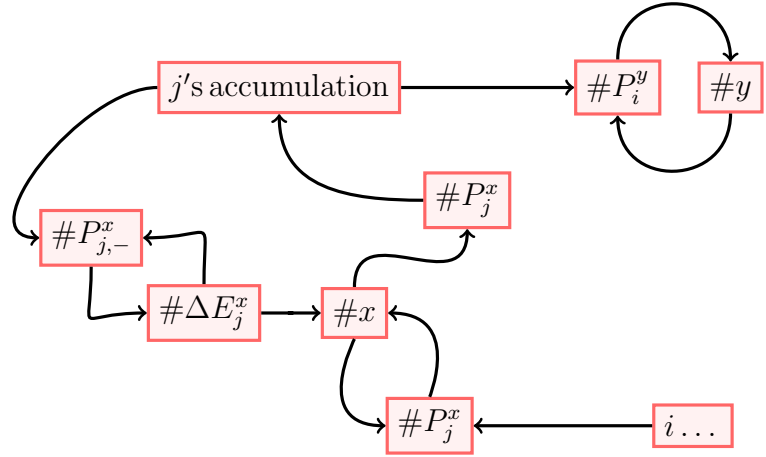
\begin{figure}[h]
\centering
\begin{tikzpicture}[
SIR/.style={rectangle, draw=red!60, fill=red!5, very thick, minimum size=5mm},]

\node[SIR]    (j)  at(3.5,0) {$j'\mathrm{s\, accumulation}$};
\node[SIR]    (P0) at(1,-2)  {$\#P_{j,-}^x$};
\node[SIR]    (E) at(2.5,-3)  {$\#\Delta E_j^x$};
\node[SIR]    (produce)  at(4.5,-3) {$\#x$};
    \node[SIR]    (trade) at(5.5,-4.5)  {$\#P_j^x$};
\node[SIR]    (revenue) at(6,-1.5)  {$\#P_j^x$};

\node[SIR]    (cash)  at(8,0)  {$\#P_i^y$};
\node[SIR]    (purchase) at(9.5,0) {$\#y$};

\node[SIR]   (i) at(9,-4.5) {$i\dots$};

\draw[->, very thick] (j.west) .. controls  +(left:1) and +(left:1)   .. (P0.west);
\draw[->, very thick] (P0.south) .. controls  +(down:1) and +(left:1)   .. (E.west);
\draw[->, very thick] (E.east) .. controls  +(right:1) and +(left:1)   .. (produce.west);
\draw[->, very thick] (E.north) .. controls  +(up:1) and +(right:1)   .. (P0.east);
\draw[->, very thick] (produce.south) .. controls  +(left:0) and +(left:1)   .. (trade.west);
\draw[->, very thick] (trade.north) .. controls  +(left:0) and +(right:1)   .. (produce.east);
\draw[->, very thick] (produce.north) .. controls  +(up:1) and +(down:1)   .. (revenue.south);
\draw[->, very thick] (revenue.west) .. controls  +(left:1) and +(down:1)   .. (j.south);

\draw[->, very thick] (j.east) .. controls  +(left:0) and +(right:0)   .. (cash.west);
\draw[->, very thick] (cash.north) .. controls  +(up:1) and +(up:1)   .. (purchase.north);
\draw[->, very thick] (purchase.south) .. controls  +(down:1) and +(down:1)   .. (cash.south);

\draw[->, very thick] (i.west) .. controls  +(left:1) and +(right:1)   .. (trade.east);

\end{tikzpicture}
\caption{A section of ecosystem showing player $j$ embedded in the network through his decisions to buy, say $y$, or execute job(s), say $x$.}
\label{Ecosystem}
\end{figure}

The flow diagram \ref{individual} shows an individual's decision-making process, and the flow diagram \ref{Ecosystem} shows a small intersection of the trades in the ecosystem, summarising the aforediscussed dynamics. In the current section, I present the set of axiomatic statements that give rise to the classical economic dynamics described by \citet{ricardo1817}. Then I examine the consequences of the absence of conserved quantities from the economy.

\vspace{1cm}

\subsection*{Axioms}

\begin{enumerate}
    \item There exist $players$ in the market, labelled as $i$, $j$, $k$ ..., which are entities with interests and the ability to carry out processes, characterised by efficiency vectors \footnote{Discussion in Sec. \ref{Phenomenology} suggests characterising each individual by their efficiencies. And since the efficiency of a given individual, in general, is job-specific (See Sec \ref{Formalism}), it can be represented by a vector quantity, $\vec \epsilon$, which I use for the abstraction of the process of trade.} (See Eq. \ref{eq:efficiency vectors}). \label{axiom:players}
    \item There exist processes, called $Jobs$, labelled as $x$, $y$, $z$ ..., which players can carry. out and owning their outcomes are the interests of the players. \label{axiom:Jobs}
    {\item Execution of each job by a player comes at his expense of a zero-sum parameter (or a conserved quantity \footnote{To repeat, a \textit{zero-sum parameter}, a quantity for which system is closed, a \textit{invariant quantity}, or a \textit{conserved quantity} all mean the same thing (a quantity/parameter whose amount/value remains unchanged in the system overall) and will be used interchangeably throughout the paper. The term \textit{Zero-sum} is commonly used in economics, whereas \textit{closed systems} and \textit{Conserved quantity} are common to Physics, and the \textit{invariant quantity} is most typical in Mathematics.}), labelled as $E$. \label{axiom:E-cost}}
    \item The outcomes of the job originally belong to the player who executed the job and can be exchanged ($traded$) between the players. \label{axiom:tradepermission}
    \item Each player makes a choice for jobs to execute and trades to engage in to maximise its own returns by either trading or executing Jobs. And the system settles into a configuration such that net energy expenditure in the economy is minimised. \label{axiom:Profit-Maximisation}
\end{enumerate}

To elaborate, a job is simply the description of the job. Abstractly speaking, arbitrary requirements can be added to the job description, and the job is considered done only when these requirements are met\footnote{Note, the most general case, where you take into account not only the net energy consumption of the processes (or, equivalently, their efficiencies), but also the time rate of the job completion, is not explicitely done here. 
However, we could consider including the time rate (instantaneous or average) as a part of the job definition, without loss of generality.}. 

Here, axioms \ref{axiom:players}, \ref{axiom:Jobs}, \ref{axiom:tradepermission}, and \ref{axiom:Profit-Maximisation} come from the Ricardian description of economics. This section will show the necessity of axiom \ref{axiom:E-cost} as necessary for the existence of trade.

\subsection*{Proof}

Axiom \ref{axiom:E-cost} says that prices have a dependence on expenditure of this zero-sum parameter, known as energy $E$. This can be shown explicitly as

\begin{equation}
\mathcal{P}_{i,-}^x=\mathcal{P}_{i,-}^x(\Delta E_i^x, \Delta E_j^y, \dots)
\label{eq:price_energy_dependence}
\end{equation}

 where $\mathcal{P}_{i,-}^x$ is the the lowest profitable selling price player $i$ can sell $x$ at, and $\Delta E_i^x$ is the energy cost of individual $i$, in executing job $x$. This expression explicitly shows the dependence of $P$, the price, on a conserved quantity, $E$, which we give the name \textit{Energy}, and also shows explicitly that the dependence on $i$ and $x$ of $\mathcal{P}$ comes from the dependence of $\Delta E$ on $i$ and $x$.



An economic system consists of players who can execute jobs and a set of jobs available to be executed. Consider each player is characterised by their efficiency vector for different jobs (collections of such vectors for all the players characterise the entire economy system), such that,

\begin{equation}
\vec{\varepsilon_i}=\begin{pmatrix} \varepsilon_i^x \\ \varepsilon_i^y \\ \vdots \end{pmatrix} \mathrm{,} \,\, \vec{\varepsilon_j}=\begin{pmatrix} \varepsilon_j^x \\ \varepsilon_j^y \\ \vdots \end{pmatrix}
\mathrm{, \dots}
\label{eq:efficiency vectors}
\end{equation}

and 

\begin{equation}
\varepsilon_i^x \xmapsto{\text{maps to}} \Delta E_i^x
\end{equation}
Let the net energy expenditure, which is the sum of energy expenditure of each individual, in the system for a particular way of distributing the jobs among the individual players, be given by 

\begin{equation}
\Delta E_{net}=\Delta E_i^x + \Delta E_j^y + \dots
\label{eq:net_E}
\end{equation}
Since this is a function of how the individual jobs are distributed amongst players, we can explicitly state $\Delta E_{net}$ as a function of this permutation. 

\begin{equation}
\Delta E'_{net}(\mathrm{Pr_a})=\Delta {E'}_{i}^{x} + \Delta {E'}_{j}^{y} + \dots
\label{eq:net_E_permutation} \mathrm{,}
\end{equation}
where primes $'$ on the variables indicate that it may not be the most efficient assignment the system can form. Here $\Pr_a$ represents a particular permutation in which the system exists. Subscript $a$ takes a natural number and corresponds to different permutations of job and player matching. $\Delta E$s on the RHS do not need this explicit representation of dependence on permutation since they already are shown in the form combination of $i$ with $x$, $j$ with $y$, and so on... 

Let's say $\Pr_1$ happens to be the ideal distribution of jobs amongst the players; i.e such that the total energy expense in the ecosystem is minimum. Such a system may not result in the most profit-maximising situation for everyone. Despite the individual's aiming to minimise their energy expenditure on the tasks and maximise savings of the would-have-been consumed energy of the system, one who saves it the most for the entire economic ecosystem generally wins, and results in the job distribution being minimal for the system. Thus, the system will tend to settle into a permutation that minimises the total $\Delta E$; the condition is represented by

\begin{equation}
\frac{\partial \Delta E_{net}}{\partial \mathrm{Pr}}\stackrel{?}{=}0
\label{eq:E_minimisation} 
\footnote{This is not an equation but merely a representation of the condition. This can be evaluated as a discretised derivative, and the Pr corresponding to the smallest value of the derivative expression is akin to a stationary point in $E$ vs $\mathrm{Pr}$ plot.}
\end{equation}

{
\begin{figure}
\centering
\begin{tikzpicture}[
        hatch distance/.store in=\hatchdistance,
        hatch distance=10pt,
        hatch thickness/.store in=\hatchthickness,
        hatch thickness=2pt
    ]
    \makeatletter
    \pgfdeclarepatternformonly[\hatchdistance,\hatchthickness]{flexible hatch}
    {\pgfqpoint{0pt}{0pt}}
    {\pgfqpoint{\hatchdistance}{\hatchdistance}}
    {\pgfpoint{\hatchdistance-1pt}{\hatchdistance-1pt}}%
    {
        \pgfsetcolor{\tikz@pattern@color}
        \pgfsetlinewidth{\hatchthickness}
        \pgfpathmoveto{\pgfqpoint{0pt}{0pt}}
        \pgfpathlineto{\pgfqpoint{\hatchdistance}{\hatchdistance}}
        \pgfusepath{stroke}
    }

    \begin{axis}[
        xlabel=$\mathcal{P}$,
        ylabel=$n(\mathcal{P})$,
        width=17cm, 
        height=13cm, 
        unit vector ratio=1 1 1, 
        xtick distance=1,
        ytick distance=1,
        xmin=0, xmax=17, 
        ymin=0, ymax=13,
        extra x ticks={6.5}, 
        extra x tick labels={\color{green} $\mathcal{P}_j$}, 
        legend pos=north east 
    ]
    \addplot[smooth, domain=0:1,
        samples=100, mark= , name path=mycurve] coordinates 
    {(0,4.3) (3,4) (5.7,6.8) (6.5,7) (7,6.8) (8.4,6.5) (9.9,4.9) (13,7.7) (15,4) (17,3.3)};
     \addlegendentry{$n(\mathcal{P})$}

    \addplot+[smooth, mark=none,
        domain=-2:-0.5,
        samples=100,
        pattern=flexible hatch,
        hatch distance=10pt,
        hatch thickness=0.5pt,
        draw=gray,
        pattern color=green,
        area legend] coordinates {(6.5,7) (7,6.8) (8.4,6.5) (9.9,4.9) (13,7.7) (15,4) (17,3.3)} \closedcycle;
        \addlegendentry{$B(n(\mathcal{P}),\mathcal{P}_j)$}
    
    
    \addplot[thick, dashed, green] coordinates {(6.5,0)(6.5,13)};
    \end{axis}
\end{tikzpicture}
\caption{Population distribution plot showing buying affordability of individuals for a given seller $j$ of produce of some job $x$.}
\label{fig:Population distribution and buying affordability}
\end{figure}
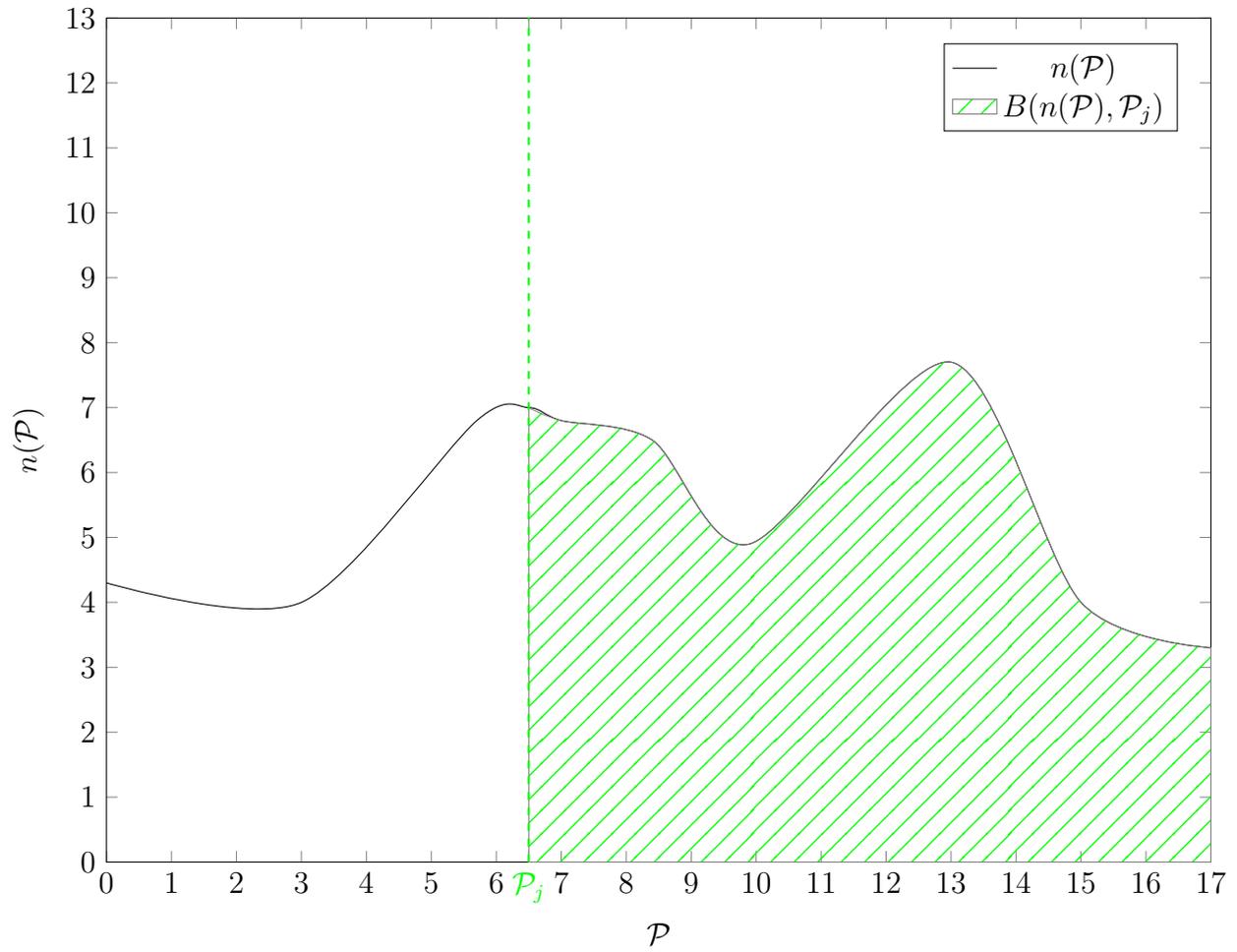
}

To analyse the maximisation of gains at the level of an individual player, consider the population distribution given in the plot in fig \ref{fig:Population distribution and buying affordability}. Here $n({\mathcal{P}})$ gives the number of players from the ecosystem who can execute a job, say $x$ (dropped through the rest of the section for simplicity), at price $\mathcal{P}$. To be accurate, $n({\mathcal{P}})$ is player density at each $\mathcal{P}$, rather than just number player; the integral of $n({\mathcal{P}})$ from $0$ to some price $\mathcal{P_j}$ will give the number of people [who prefer to trade].  Thus, by construction, since the plot is of the entire population participating in the economic game,

\begin{equation}
\int_0^{\p_{max}}n(\mathcal{P}).d\mathcal{P}=N
\end{equation}
must always be true, where $N$ is total number of players in the system, $\p_{max}$ is the cost of most inefficient buyer on $n(\p)$ curve. 

Naturally, if a player in the ecosystem finds a seller who can sell the produce of the same job lower price than his own cost of production, a trade is preferred over the self-execution of the job\footnote{In real systems, the information flow is not instantaneous, and not entire ecosystem will be taking part in evaluation of the worth of trading with $j$. 
The complications, however, merely change the number of participants evaluating $j$'s produce, which the proof does not hinge upon. Thus, I only consider the simple and natural model where an individual compares the$j$'s cost of production to their self-production cost.}. Say a player $j$ in market prices his produce at a price $\mathcal{P}_j$. The integral of $n({\mathcal{P}})$ from $0$ to $\mathcal{P}_j$ gives the number of players not entering into a trade with $j$. $N$ minus this quantity gives the number of players entering into a trade with $j$.  i.e.

\begin{equation}
\int_{\mathcal{P}_j}^{\p_{max}} n(\mathcal{P}).d\mathcal{P}=B(\mathcal{P}_j)
\label{eq:integral giving total number of buyers}
\end{equation}



or

\begin{equation}
    B(\mathcal{P}_j)=
    N-\int_{0}^{\p_j}n(\mathcal{P}).d\mathcal{P}
\end{equation}
Here, $B(\mathcal{P}_j)$ is the number of buyers from he ecosystem who would be willing to buy at the price attempted by $j$, considering each player's self-cost of doing the job. $B$ is also a function of the function $n(\p)$, but will be dropped throughout the section for compactness and to avoid unhelpful distraction. At the individual level, since each has a different $\Delta E$, giving them a different price margin, the maximising condition on gains is

\begin{equation}
\frac{\partial[(\mathcal{P}_j-\mathcal{P}_{j,-})B(\mathcal{P}_j)]}{\partial \mathcal{P}_j}=0 
\footnote{One could equivalently use quantities expressed as a function of $\Delta E$'s instead of $\mathcal{P}$, since there exists a map between $\Delta E$'s and $\p$'s. (see eq. \ref{eq:price_energy_dependence})}
\mathrm{,} 
\end{equation}
where $\mathcal{P}_{j,-}$ is the least price $j$ could sell at, and $\mathcal{P}_{j}$ is the actual price at which he attempts to sell. $B(\mathcal{P}_j)$ gives the number of buyers from he ecosystem who would be willing to buy at the price attempted by $j$, considering each player's self-cost of doing the job. A generic distribution as follows could be inserted in the expression here in place of $n(\mathcal{P})$ or $B(\mathcal{P}_j)$ to analyse how the characteristics of the ecosystem with a particular arbitrary distribution of individuals' efficiencies--- or a Pareto distribution to depict the reality.

{
Now, let us assume the converse of the claim made in this paper: that the phenomena of trade can exist without the existence of a conserved quantity (or a zero-sum parameter).\label{counterclaim}} When the execution of a job is not associated with the expense of a conserved quantity, or in other words, creation of energy (the quantity associated with the job execution) is allowed, each individual's cost of creation goes to $0$. Thus, the distribution $n(P)$ would take the form:

\begin{equation}
n(\mathcal{P})=\delta(\mathcal{P}) \equiv \begin{cases}
    N, & \text{if } \p = 0 \\
    0, & \text{if } \p \neq 0
\end{cases}
\label{eq:delta_distribution}
\end{equation}
The integral over $n(\mathcal{P})$ giving the total number of buyers $B(\mathcal{P}_j)$ is

\begin{align}
    B(\mathcal{P}_j)=\int_{\p_j}^{\p_{max}}n(\mathcal{P}).d\mathcal{P} & = \int_{0}^{0}n(\p).d\p = 0
\label{eq:zero buyers}
\end{align}
On the other hand, 

\begin{equation}
    \int_0^{\p_{max}}n(\mathcal{P}).d\mathcal{P}=N \mathrm{,}
\end{equation}
which is a contradiction. Alternatively, from eq. \ref{eq:delta_distribution}, $n(\mathcal{P})$ curve is zero everywhere except at $\mathcal{P}=0$, the upper limit of the above integral can be moved to $0$ (or to any other number between $0$ and $\infty$). Thus.

\begin{equation}
    \int_0^{\infty}n(\mathcal{P}).d\mathcal{P}=\int_0^{0}n(\mathcal{P}).d\mathcal{P}=N \mathrm{.}
\end{equation}

This contradicts eq. \ref{eq:zero buyers}. Therefore, the absence of axiom \ref{axiom:E-cost} leads to a illy defined theory, and is a necessary statement for axiomatic completeness of the economic theories, displaying the need for conserved quantiteis (or zero-sum parameters). This demonstrates that the economy cannot exist in the form we see it without the existence of conserved physical quantities. Fortunately, we know from Classical Physics that all the dynamical events in the universe adhere to the conservation laws for two quantities: $Energy$ and $momentum$ \cite{Newton’sWikipedia_contributors_2025}. While momentum is conserved only when no amount of potential energy is being converted to kinetic, Energy by itself, which is the sum of both potential and kinetic, is always conserved. Thus, the zero-sum parameter associated with job execution, which I named $Energy$ (see Axiom \ref{axiom:E-cost}), must clearly be the same conserved quantity described in Physics as the $Energy$.

Note a less interesting result, that the $n(\mathcal{P})$ curve also takes delta form (see eq. \ref{eq:delta_distribution}) when efficiencies of all the individuals are the same for a given job $x$. However, as stated at the beginning of the section, the individuals are characterised by their efficiencies (\ref{eq:efficiency vectors}), which is to say that if two individuals have the same efficiency, there exists no factor allowing us to distinguish those two, and there is no meaningful way for trade to exist between the two. However, for completeness, it is worth mentioning either of the statements as necessary for the existence of trade: 1) Existence of a zero-sum parameter (or a conserved quantity) that $Job$ executions cost, and 2) the ability to distinguish the players in the ecosystem through their efficiency vectors.

\section{Energy is not wealth, nevertheless.}
\label{sec:corollaries:Energy is not wealth though.}

\textbf{It is to arrive at the best solution to the optimisation problem designed to maximise the margin. The Input to this function is—EVERYTHING—you can think of, literally.} The economy is, therefore, based on the exploitation of differences in efficiencies. But not only do people differ in their efficiencies, they also differ in their ability to judge. They differ in their ability to accurately determine an exhaustive sequence of actions that would maximise their margin. Those good at it know better what to trade for and know exactly when to†\textsuperscript{6}. As a consequence, individuals around the globe differ in their wealth. You cannot create energy\textsuperscript{3} \cite{Newton’sWikipedia_contributors_2025}. And precisely because you cannot create energy, it always costs you to produce something, anything. You have to expend from what you already have before you can have something to sell and hope that it would be deemed valuable by others: that they find they would be able to increase their overall margin by having what you produce after accounting for the price initially paid; and that the margin is maximum when they buy it from you rather than anyone else—even themselves.

\section{Energy is the zero-sum parameter, not wealth.}
\label{sec:corollaries:Energy is the zero-sum parameter, not wealth.}

\textbf{"The rich get richer and the poor get poorer"; not quite.} Now, this might seem to support the familiar rhetoric "The rich get richer and the poor get poorer". However, it doesn't. 

Engaging in trade maximises everyone's returns. Trade enables everyone to live more efficiently than otherwise. It makes everyone rich. It’s only that the rates of wealth gain differ. Some gain wealth faster than others. (This is not to say that there aren't exceptions; that there ain't any who fall back, who get poorer. But these are exceptions, not the norm.)

This explains the Pareto distribution of wealth \textsuperscript{5}. If someone is better (more efficient) at producing what you produce, they would be chosen to buy it from. Not only shall he have a wider margin due to his efficiency, but also all the customers. The business shall be his, and the wealth that comes along. In other words, those most efficient at catering to demand get to cater to the most demand. They produce more because they are best at producing it. They are chosen to consume more (to produce it) because they consume it the most efficiently.

The currency system, instead of causing, only makes previously unnoticeable phenomenon—the Pareto distribution of wealth—more visible. With it, we don't necessarily need to be good at things mutually needed. There is a universal way to measure wealth (which is a measure of energy[3] saved): the commonly agreed-upon currency we use to exchange with all other goods and/or services. The amount of money possessed by an individual is an indicator of how efficient he has been (on average) heretofore\textsuperscript{1}, or, put differently, how much would-have-been-consumed energy he had saved of those he traded with.

\section{Deviation from perfect trade and execution decisions}
\label{sec:Errenous estimates}
\subsection*{Errenous estimates}

Now, an objection to this hypothesis, which suggests that the pricing of goods/services is based on the potential energy consumption (would-have-been-consumed) avoided, rest same, by the buyers (the upper bound of price), could be the fact that there are things whose price couldn’t be accounted for based on the saving of the would-have-been-consumed Energy: gold, pearls, diamonds. Here, the prices are dictated by rarity: a fault in estimation. But, looking at it carefully, one can see that these too fail to defy the energy-saving rule. It's not based on rarity merely—with no concern for the zero-sum parameter like Energy. This doesn't [always] lead to a decrement in efficiency\textsuperscript{1}. Buying the rare isn't [always] equal to squandering the money\textsuperscript{1}.

\textbf{The key is the consistency of the fault:} If an item is deemed valuable universally and consistently, irrespective of its practical usefulness (saving of the would-have-been-consumed energy)—or, more accurately, practical usefulness not commensurate with its price—it would continue to be deemed valuable. The diamond I bought (with money) behaves less like a good and more like money itself. Like paper notes, it can always be exchanged for something else (perhaps intrinsically valuable). The efficiency isn't lost, perhaps it's gained. The energy isn't wasted; perhaps it's saved in the overall system.

This doesn't weaken the hypothesis for the very reason that currencies in the form of precious metals, like gold and silver, might sometimes be unreliable, because if the value printed on the tiny plate of gold is less than the value of gold required to make it, it is more profitable to sell it as gold rather than currency. A small piece of paper can hardly be worth \rupee100. Thus, if we all agreed to mark it at \rupee100, and agreed to use it as such, it is self-impoverishing to sell it as a piece of paper, for its practical usefulness. (One can easily replace something of practical utility in these examples: perhaps a limited edition car). So, if the true value of an object dictated by its functional utility is more than its value dictated by rarity—in other words, if it’s not rare enough—the former would be the accepted value. The existence of cases of the latter type could be attributed to our inability to judge what is of true value. It is a fault in our ability to judge, in our instinct.

However, not all faults are equally faulty. If a particular type of fault makes us value things not intrinsically valuable, or “unproductive” assets, and if that fault is consistent across time (meaning it doesn’t get corrected), then one can still expect that thing to be considered valuable in future. And of course, if the fault is valuing the rare, rarity too has to be an undying one. Therefore, and more generally, a fault that is consistent isn’t as faulty.\footnote{This specific fault, valuing the rare, perhaps evolved due to the advantage statistically it might have provided. If something is rare, it might sometimes be worth getting ownership over it before being certain of its true value. That which is rare gives less window of time to judge its value and own it. The strategy might have allowed individuals to catch valuable deals, valuable enough and often enough.}



$$$$




\section{Summary}

What does this imply for the comparative advantage theory and other results based on it? With these statements, the theory of comparative advantage results in vague, incompletely defined dynamics. If people can create energy, they can get infinitely rich (for free). They can sell energy to get there. There is no upper bound on prices. Anyone can buy anything—for free (because you got energy for free). But, as they can create energy, they wouldn't have to buy it (or anything) from each other. They can produce everything they need themselves, too (again, for free)\textsuperscript{3}[3]. There is no lower bound on prices since the product came into existence with zero expense. But, also because everyone can create energy, they can feed themselves (by definition) and be self-sustained—there is no need to get rich. Thus, not only is the basis for when to engage in trade absent, but the utility of trade is so too. The trade IS because unlimited energy ISN'T.

Thus, in a universe where energy creation isn’t forbidden, trade doesn’t exist (assuming the entities analogous to traders (people) exist).

\section{Future scope}
\subsection*{Modelling errors in estimates as biased random walk}
Consider the vector 

\begin{equation}
\vec{\mathcal{P}_{j}}^x 
= \begin{pmatrix} \mathcal{P}_{i-j}^x \\ \mathcal{P}_{j-j}^x \\ \mathcal{P}_{3-j}^x \\ \vdots \end{pmatrix} 
=\begin{pmatrix} \mathcal{P}_{1-2}^x \\ 0 \\ \mathcal{P}_{3-2}^x \\ \vdots \end{pmatrix}
\end{equation}
representing prices that player $j$ can sell $x$ at to each individual in the market.

Profit maximisation optimisation could be done across the elements of this matrix to find the optimal pricing, considering all players being all-knowing \footnote{This is not the same as completely rational players in that rational players often gather information across time as they play the game more, whereas the all-knowing players simply know, with no need for finding more} as buyers. However, this is less representative of the real-world applications. Instead of an all-knowing player in the real market, we have those who are trying to be all-knowing, but only manage to be approximately so. Thus, we can consider a spread around the true values arrived at by erroneous evaluation by the individuals in the market. It is natural to choose a Gaussian kind of distribution spread around the true value as it has two properties that could depict the real-world scenario: 1) the farther a value from the true value you consider along the number line, the less likely an individual to estimate it to be the true value, and thus, the fewer the individuals evaluating the product, say $x$, at that price; and 2) Most people are likely to be centered around the true value, while none may actually hit the exact true value precisely.  Thus, each of the single-valued elements of this vector can be replaced by probability density functions [denoting the fraction of the population evaluating the true value of a trade to be $\mathcal{P}$], which could in turn be related to the probability of a purchase happening at that price. 

Further, each individual's evaluation could be modelled to experience a biased random walk across time, centred around the true price. This lets us find the expectation value of price, $\langle\Delta\mathcal{P}^x_{i-j}\rangle$. Random walk description allows quantities analogous to temperature, pressure, and entropy could be found.


\section{Conclusion}
In this paper, I put the Ricardian description of the economy on an axiomatic footing in sec. \ref{Phenomenology}. Next, I demonstrate the axiomatic incompleteness of the set of axioms obtained merely from the Ricardian description. Through this, I show the mathematical necessity of an axiomatic statement binding economic process with conserved quantities, demonstrating that trade cannot exist in the absence of an associated conserved quantity with all the economic processes (see sec. \ref{Formalism}). Albeit a less interesting result, but on equal axiomatic footing, I note that trade cannot exist if players participating in an economic system have the same efficiency vectors (see \ref{eq:efficiency vectors}). 

Further, I discuss how the zero-sum parameter in economics is not the wealth but the energy, and introduce the concept of \textit{would-have-been-consumed energy}, giving a physical basis for wealth accumulation. Lastly, I discuss the deviation from the ideal decision between trade and job execution, and pave the way for future work on modelling such errors in the estimation of the true value of the \textit{jobs} through a random walk biased towards the true value.

\section*{Acknowledgments}

I'd like to thank Barry Goodwin for his valuable time and for providing input on improving this paper, as well as validating the accuracy of various tenets of economics used. 

\clearpage
\bibliographystyle{plainnat}
\bibliography{sample}

@misc{comparative-advantage,
    title = {On the {Principles} of {Political} {Economy} and {Taxation}},
    url = {https://www.econlib.org/library/Ricardo/ricP.html},
    language = {en-US},
    year = {2025},
    urldate = {2025-09-03},
    journal = {Econlib},
    author = {Ricardo David},
}

@article{soo2011gains,
  title={The gains from specialisation and comparative advantage},
  author={Soo, Kwok Tong},
  journal={Lancaster University},
  year={2011}
}

@article{Behavioral_economics_pioneered_by_Daniel_Kahneman_and_Amos_Tversky,
 ISSN = {00129682, 14680262},
 URL = {http://www.jstor.org/stable/1914185},
 author = {Daniel Kahneman and Amos Tversky},
 journal = {Econometrica},
 number = {2},
 pages = {263--291},
 publisher = {[Wiley, Econometric Society]},
 title = {Prospect Theory: An Analysis of Decision under Risk},
 urldate = {2025-10-15},
 volume = {47},
 year = {1979},
}

@book{Information_economics_pioneered_b_George_Akerlof_Michael_Spence_and_Joseph_Stiglitz_explores_the_role_of_information_asymmetry_in_economic_transactions,
 ISBN = {9780691130613},
 URL = {http://www.jstor.org/stable/j.ctt1r2gkx},
 author = {John von Neumann and Oskar Morgenstern and Ariel Rubinstein},
 publisher = {Princeton University Press},
 title = {Theory of Games and Economic Behavior (60th Anniversary Commemorative Edition)},
 urldate = {2025-10-15},
 year = {1944},
}

@misc{Newton’sWikipedia_contributors_2025, 
title={Newton’s laws of motion}, 
url={https://en.wikipedia.org/wiki/Newton%27s_laws_of_motion#Prerequisites}, 
journal={Wikipedia}, author={Wikipedia contributors}, year={2025}, 
month=sep, 
year = {2025},
}

@book{ridley2010rational,
  title={The Rational Optimist: How Prosperity Evolves},
  author={Ridley, Matt},
  year={2010},
  publisher={Harper},
  address={New York},
  isbn={9780061452055},
}

@book{ricardo1817,
  title = {On the principles of political economy and taxation},
  author = {David Ricardo},
  publisher = {London : John Murray, 1817.},
  year = {1817},
  url = {https://search.library.wisc.edu/catalog/999736985502121}
}

@book{smith1776,
  author    = {Smith, Adam},
  title     = {An Inquiry into the Nature and Causes of the Wealth of Nations},
  publisher = {W. Strahan and T. Cadell},
  year      = {1776},
  address   = {London}
}

\end{document}